# ON THE CHALLENGES AHEAD OF SPATIAL SCIENTOMETRICS FOCUSING ON THE CITY LEVEL

**György Csomós**
University of Debrecen Department of Civil Engineering
csomos@eng.unideb.hu

**Abstract**

Since the mid-1970s, it has become highly acknowledged to measure and evaluate changes in international research collaborations and the scientific performance of institutions and countries through the prism of bibliometric and scientometric data. Spatial bibliometrics and scientometrics (henceforward spatial scientometrics) have traditionally focused on examining both country and regional levels; however, in recent years, numerous spatial analyses on the city level have been carried out. While city-level scientometric analyses have gained popularity among policymakers and statistical/economic research organizations, researchers in the field of bibliometrics are divided regarding whether it is possible to observe the spatial unit 'city' through bibliometric and scientometric tools. After systematically scrutinizing relevant studies in the field, three major problems have been identified: 1) there is no standardized method of how cities should be defined and how metropolitan areas should be delineated, 2) there is no standardized method of how bibliometric and scientometric data on the city level should be collected and processed and 3) it is not clearly defined how cities can profit from the results of bibliometric and scientometric analysis focusing on them. This paper investigates major challenges ahead of spatial scientometrics, focusing on the city level and presents some possible solutions.

## 1 Introduction

As globalization has intensified over the last five decades, large global systems such as trade and manufacturing have become even more integrated. International science has also witnessed deep integration (Gazni et al. 2012; Georghiou 1998). Due to the rapid development of information and communication technologies (primarily that of the Internet), and cheap air travel, physical barriers in the way of international research collaboration have been removed to a significant degree (Hoekman 2012; Pan et al. 2012). The geographical pattern of international research collaboration has significantly changed over time and newly emerging countries (China in the first place) have positioned themselves at the focal points of global science. Measuring and evaluating changes in international research collaboration and the scientific performance of institutions and countries through the prism of bibliometric and scientometric data have become key. The examination of the spatial aspect of science on the basis of bibliometric and scientometric data is in the scope of spatial scientometrics, a quantitative research field being positioned between the fields of scientometrics and geography (i.e., the geography of science) (Frenken et al. 2009a; Gao et al. 2013; Hoekman 2012). Parallel with the globalization of science, spatial scientometric studies have become more common. As demonstrated by Frenken et al. (2009a), spatial scientometrics initially involved the country level in the research focus — the works performed by Narin and Carpenter (1975) and Frame et al. (1977) are considered particularly important in the field — and the majority of spatial scientometric studies have still been focusing on examining the country level (see, for example, Almeida et al. 2009; Bornmann et al. 2018; Gazni et al. 2012;

Glänzel 2001; Horta and Veloso 2007; Lee et al. 2011; Leydesdorff et al. 2014; Miao et al. 2018; OECD and CSIC 2016; Stephan and Levin 2001; Wang et al. 2017). Besides country-level scientometric analyses, based on experiences gained from them, many spatial scientometric studies have been carried out with sublevel territorial units (e.g., federal states, provinces, and NUTS regions[1]) in focus. This spatial scientometric approach has become widespread in countries with a hierarchized spatial structure and where the sublevel spatial division is endowed with a strong administrative function (see, for example, the case of Canada: Godin and Ippersiel 1996; China: He et al. 2005; Liu et al. 2015; Zhou et al. 2009; Italy: Abramo et al. 2014; Abramo and D'Angelo 2015; the Netherlands: Ponds et al. 2007; and the United States: Carvalho and Batty 2006; Hohmann et al. 2018; Kamalski and Plume 2013; Thompson 2018). Some studies go beyond focusing on formal regions (sublevel spatial divisions) and examine the scientometric characteristics of city-regions that were created for special analytical purposes (see, for example, Bornmann and Waltman 2011; Grossetti et al. 2014; Maisonobe et al. 2018b, 2019; Matthiessen and Schwarz 1999; Matthiessen et al. 2002). Since the mid-1960s, urban geographers and regional economists have been paying high attention to cities with a special role in the global economy (labelling them as world/global cities) and the networks being created by them (Alderson and Beckfield 2004; Borchert 1978; Cohen 1981; Friedmann 1986; Godfrey and Zhou 1999; Hall 1966; Heenan 1977; Hymer 1972; Sassen 1991; Taylor 2001; Wheeler 1985), while spatial scientometric research focusing on the city level has only been given impetus since the beginning of the 2000s, after the publication of Matthiessen & Schwarz's (1999) pioneering work. However, in recent years, spatial scientometric studies examining the position of cities in global science have become widespread. A significant part of global research activity has traditionally been concentrated in certain cities and their metropolitan hinterlands (Van Noorden 2010). These cities are considered to be the most powerful magnets for attracting members of the creative class (e.g., scientists, researchers, and engineers) and innovative companies of which combination with a large amount of research funds available will make these cities the most significant location of new knowledge production and international centres of innovation. This process can be characterized by a self-reinforcing upward spiral. In recent years, the global science system has become more fragmented and diverse, due to the fact that powerful emerging cities have appeared in the vertices (Csomós 2018a; Maisonobe et al. 2018a; Nature Index 2018), forcing the traditional scientific centres to compete with them (to acquire globally available research funds, for instance). This new approach within the frame of spatial scientometrics focuses on how cities participate in global science. Some relevant works are listed below:

- The paper authored by Matthiessen and Schwarz (1999) presents an analysis of scientific strength by output produced by authors from the 'greater' urban regions of Europe.
- Bornmann et al. (2011) and Bornmann and Leydesdorff (2011, 2012) identify and map cities which are considered to be centres of excellence in scientific research on the basis of the size and frequency of the production of highly cited papers.
- Bornmann and de Moya-Anegón (2018) map German cities, with most papers belonging to the 1% most frequently cited papers, within their subject area and publication year. Bornmann and de Moya-Anegón (2019) detect hot and cold spots in the United States based on bibliomteric data produced by institutions.
- Maisonobe et al. (2016, 2017) investigate cities' publication output and collaboration network from different aspects while Grossetti et al. (2014) examine the global and national deconcentration of scientific activities through the prism of cities.
- Csomós and Tóth (2016) explore the spatial distribution of scientific publications created by multinational corporations from two geographical approaches. Csomós (2018a) examines the publication dynamics, collaboration pattern and disciplinary profile of more than 2,000 cities worldwide while Csomós (2018b) reveals factors that may influence cities' publication efficiency (i.e., the highly cited paper ratio of publications produced in cities). Csomós and Lengyel (2019) visualize the efficiency of international scientific collaboration of city-dyads.
- Leydesdorff and Persson (2010) display co-authorship, collaboration networks between cities by using mapping and network visualization programs.
- The paper authored by Wu (2013) proposes a citation rank based on spatial diversity in terms of cities and countries, focusing on the measurement of the spatial aspect in citation networks.
- Jiang et al. (2017) investigate the spatial patterns of R&D collaborations of Chinese cities by using co-patent data.
- Andersson et al. (2014) reveal the internal spatial structure of China's scientific output while Ma et al. (2014) investigate the internal collaboration network of Chinese cities in terms of geographical proximity.
- Catini et al. (2015) explore spatially concentrated innovation clusters within metropolitan areas by geocoding publication data.

---

[1] NUTS is the abbreviation of the French term 'Nomenclature des unités territoriales statistiques (Nomenclature of Territorial Units for Statistics)' a geocode standard for referencing the subdivisions of the EU member states for statistical purposes.

The credibility of the scientific outcomes of the studies listed above is only acceptable if we are aware of the general limitations of scientometrics focusing on the city level. Some of these limitations have been well acknowledged by urban geographers and urban planners for a long time, and efforts have been made to eliminate them. In the following section, major challenges ahead of spatial scientometrics focusing on the city level and their possible solutions will be demonstrated.

## 2 Challenges ahead of spatial scientometrics focusing on city level

### 2.1 Challenges stemming from lack of standardized methods of defining cities and delineating metropolitan areas

In recent years, urbanization, the shift in residence of the human population from rural to urban areas, combined with the overall growth of urban population due to natural increase, has witnessed an unprecedented rate of growth. According to the United Nations, in 2018, 55 percent of the world's population lived in urban areas, a proportion that is expected to increase to 68 percent by 2050 (UN, 2018).

Settlements are of various size in terms of population and area worldwide and both their status in the national administrative system and their position in the national and international urban network are highly different as well. Due to the unprecedented population growth being experienced today, the built-up area of settlements (primarily that of large cities) is expanding continuously (as demonstrated by OECD.Stat[2], and visualized by the EC GHSL[3]), neighbouring built-up areas are becoming spatially more integrated and eventually larger settlements fully incorporate the smaller ones. In addition, thanks to the development of mobility and urban transportation infrastructure, the relative distance in terms of travel time between the central city and outer settlements is becoming increasingly smaller[4]. As a result of these processes, significantly large monocentric or (more frequently) polycentric city-regions are coming into existence with a large city or several large cities in the centre surrounded by many suburbs and rural areas of various sizes. These city-regions can host dozens, sometimes hundreds of settlements having different administrative status and size in terms of population. However, it is observed that there is a diversified but close functional connection amongst settlements primarily in terms of daily commuting of the workforce (with a dominant direction towards the central city) (Knox and McCarthy 2012). For urban geographers, urban planners and regional economists, the delineation of urban agglomerations, metropolitan areas and city-regions (similar terms but not synonyms) has been of high importance for a long time (see, for example, Bode 2008; Davoudi 2008; Florida et al. 2008; Harrison 2010; Jonas 2013; Jonas and Moisio 2018; Lang and Knox 2009; Lidström 2018; Moisio 2018; Neuman and Hull 2009; Rodríguez-Pose 2008; Roy 2009; Schnore 1962; Scott 2019), and it is a key question for statistical and economic organizations committed to collecting and analyzing spatial data as well (ESPON 2007; INSEE 2011; OECD 2012; OMB 2017; Simeonova et al. 2018). There is, however, no standardized method of how metropolitan areas should be delineated and without consensus, multiple metropolitan areas in terms of population and area can be designated surrounding the same central city. For example, the Greater Tokyo Area corresponds to at least six metropolitan areas of which area and population range from 10,000 to 37,000 square kilometres, and 35 million to 44 million inhabitants. In some giant metropolitan areas, hundreds (sometimes more than a thousand) of settlements are located. For example, the Paris metropolitan area covers almost the entire Île-de-France region with an area of 17,000 square kilometres and encompasses approximately 1,800 settlements.

---

[2] OECD.Stat, Built-up area and built-up area change in Functional Urban Areas: https://stats.oecd.org/Index.aspx?DataSetCode=BUILT_UP_FUA

[3] European Commission, Global Human Settlement Layer: https://ghsl.jrc.ec.europa.eu/visualisation.php#

[4] Since the beginning of the 1990s, the 'death of distance' thesis has gained popularity amongst geographers and economists (see, for example, Cairncross 2001; Morgan 2001; and O'Brien, 1992); however, according to Frenken et al. (2009b), 'distance' is still a significant factor influencing collaboration between researchers.

The delineation method is of significant importance when the main goal of the analysis is to measure and evaluate the quality and quantity of scientific activities being conducted in cities and/or metropolitan areas. In addition, the non-standardized delineation method of metropolitan areas is considered to be one of the major sources of the problems that spatial scientometrics are faced with. For example, researchers affiliated with Stanford University, one of the world's top universities, published 12,980 SCI/SSCI papers in 2016. Stanford University is located in Stanford, California, an area which is a census-designated place being created for statistical purposes by the United States Census Bureau, and the residential population (~14,000 people) of which is less than half of the daily population (~35,000 people). Stanford is adjacent to Palo Alto, a city that is five times the size of Stanford, and both settlements are located nearly 35 kilometres north from San Jose, a metropolis with one million inhabitants, and 60 kilometres south from San Francisco, a world city with approximately 900,000 inhabitants. San Jose and San Francisco are two of the principle cities in the San Jose–San Francisco–Oakland, CA Combined Statistical Area (which includes the less extensive San Francisco Bay Area) encompassing 12 counties with a total population of 8.8 million people. The size of this city-region is now comparable in terms of area, population or even economic output with that of Moscow, London, Paris, Sydney, Tokyo, New York and Los Angeles (Bouchet et al. 2018). However, it is still substantially challenging to make any metropolitan areas in the world comparable in terms of size with Chinese cities, particularly when bibliometrics and scientometrics are being focused upon. Chinese cities are defined in an administrative way (Swerts 2017), and they do not have metropolitan areas in the traditional sense (Fang and Yu 2017). For example, Beijing, a municipality under the direct administration of the central government, covers approximately 16,000 square kilometres with a population of more than 21 million people, and encompasses 16 districts out of which the built-up area of the city extends into six inner urban districts (Yang et al. 2013). In the suburban and rural districts of Beijing, almost 200 towns and townships, and thousands of communities and villages are located, each of them forming part of Beijing municipality. In contrast to Beijing, Boston (more precisely: the City of Boston, the locality being reported by authors on the publications as the geographical location of their affiliation) covers an area of 232 square kilometres with a population of 685,000 people, and it does not include the adjacent Cambridge, the city being home to the main campuses of Harvard University and the Massachusetts Institute of Technology (MIT). Beijing (the municipality) and (the City of) Boston are obviously not on the same tier in terms of area, population, economic power or even publication output. However, Greater Boston (being described either as the Boston-Cambridge-Newton, MA-NH Metropolitan Statistical Area with a population of 4.8 million people, or the Boston-Worcester-Providence, MA-RI-NH-CT Combined Statistical Area with 8.2 million inhabitants) is of comparable size to Beijing. In the context of spatial scientometrics, it is clear which city is understood under the name of Beijing when investigating, for example, the publication output, but which Boston should be compared with Beijing? Is Stanford an independent settlement, or should it be observed together with Palo Alto, or rather as part of the San Jose-Sunnyvale-Santa Clara, CA Metropolitan Statistical Area, or the more extensive San Jose–San Francisco–Oakland, CA Combined Statistical Area, or the San Francisco Bay Area?

Scientometric studies focusing on the city level respond to the above-described problem generated by the non-standardized delineation method of metropolitan areas in two ways: one part of such studies ignores delineating metropolitan areas and considers settlements as autonomous territorial entities, while another part of them focuses on already existing metropolitan areas or creates metropolitan areas exclusively for analytical purposes (Table 1).

Table 1. Delineation and counting methods applied in scientometric studies focusing on the city level

| Publication | Spatial unit of analysis | Method of delineation | Counting method | Area covered | Database |
|---|---|---|---|---|---|
| Bornmann and Leydesdorff (2011, 2012) | cities | cities correspond to settlements reported by the authors on the publication as the geographical location of their affiliation | integer counting | global | Web of Science |
| Grossetti et al. (2014), Maisonobe et al. (2016, 2017, 2018a, b) | already existing metropolitan areas and urban agglomerations created exclusively for analytical purposes | delineation is based on population density data in the case of large metropolitan areas and a distance threshold of 40 km is used for merging smaller urban entities | fractional counting | global | Web of Science |
| Andersson et al. (2014) | city-regions | delineation is based on labour-market areas in general and urban districts with their rural hinterlands in cases of Chinese cities | not specified but most probably integer counting | global with China in focus | Web of Science |
| Ma et al. (2014) | cities | cities correspond to administrative divisions within Chinese administrative structure | integer counting | China | Web of Science |
| Bornmann and de Moya-Anegón (2018) | cities | cities correspond to administrative divisions within German administrative structure | integrating integer-counted bibliometric data of institutions on city level | Germany | Scopus |
| Bornmann and de Moya-Anegón (2019) | cities | cities correspond to settlements reported by authors of the publication as the geographical location of their affiliation | not relevant | United States | SCImago database (that uses Scopus data) |
| Catini et al. (2015) | clusters within metropolitan areas | clusters within metropolitan areas identified by geocoding publication data | not specified | global with some major cities in focus | PubMed |
| Matthiessen and Schwarz (1999) | urban agglomerations | delineation based upon NUREC delineation concept (NUREC, 1994) with an additional delineation method by combining neighbouring agglomerations when transport time between city centres is below 45 minutes | not specified but most probably integer counting | Europe | Web of Science |

| Matthiessen et al. (2002) | urban agglomerations | delineation based upon NUREC delineation concept (NUREC, 1994) with an additional delineation method by combining neighbouring agglomerations when transport time between city centres is below 45 minutes | not specified but most probably integer counting | global | Web of Science |
|---|---|---|---|---|---|
| Leydesdorff and Persson (2010) | cities | cities correspond to settlements reported by authors of the publication as the geographical location of their affiliation | integer counting | global | Web of Science and Scopus |
| Bornmann et al. (2011) | cities | cities correspond to settlements reported by authors of the publication as the geographical location of their affiliation | integer counting | global | Scopus |
| Jiang et al. (2017) | cities | cities correspond to administrative divisions within Chinese administrative structure | not relevant (analysis based upon patent data) | China | Chinese Patent Database |
| Wu (2013) | cities | cities correspond to settlements reported by authors of the publication as the geographical location of their affiliation | not relevant (citation analysis) | global | Web of Science |
| Csomós and Tóth (2016) | already existing metropolitan areas | delineation of metropolitan areas based on information provided by national statistical organizations and the OECD | integrating integer-counted bibliometric data of companies on city level | global | Scopus |
| Csomós (2018a) | cities | cities correspond to settlements reported by authors of the publication as the geographical location of their affiliation | integer counting | global | Scopus |
| Csomós (2018b), Csomós and Lengyel (2019) | cities | cities correspond to settlements reported by authors of the publication as the geographical location of their affiliation | integer counting | global | Web of Science |
| Masselot (2016) | cities | cities correspond to settlements reported by authors of the publication as the geographical location of their affiliation | integer counting of citations | global | PubMed |

In a set of papers, Maisonobe et al. (2016, 2017, 2018b), and in a previous study, Grossetti et al. (2014), created urban agglomerations exclusively to analyze publication characteristics. In these studies, the delineation of large metropolitan areas is based on population density data; furthermore, a distance threshold of 40 km is used for merging smaller urban entities. The most straightforward method of how publishing localities should be regrouped into metropolitan areas (urban agglomerations as referred to by the authors) is applied in the above-mentioned studies, and consequently their results are considered to be the most trustworthy. Maisonobe et al. (2016: 1027) assert that their 'aim was to produce universal criteria, and not divisions corresponding to a juxtaposition of national criteria (for example, using

Metropolitan Statistical Areas (MSAs) for the USA…)'. However, the delineation method being presented in their works has some special outcomes. For example, it is highly unusual to observe Ann Arbor separately from Detroit (Figure 1 demonstrates that the built-up areas of Ann Arbor and Detroit are closely adjacent to each other), primarily in light of the fact that some polycentric metropolitan regions encompassing more than one metropolitan area (taking the San Francisco Bay Area, and the Osaka-Kyoto city-region as examples) are considered as a whole.

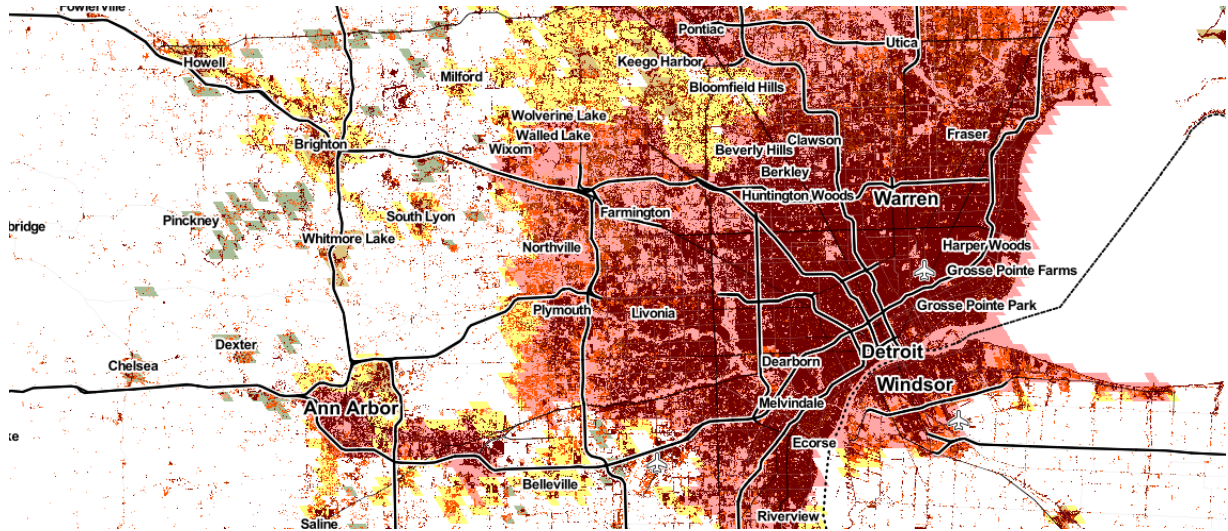

Figure 1. Adjacent built-up areas of Ann Arbor and Detroit as visualized by the EC GHSL.

Matthiessen & Schwarz (1999), and Matthiessen et al. (2002) define urban agglomerations to measure publication output, and the delineation is based upon the NUREC delineation concept (NUREC, 1994) with an additional delineation method by combining neighbouring agglomerations when transport time between city centres is below 45 minutes. Naturally, the delineation method applied in these studies may be considered appropriate in the context of spatial scientometrics; however, for urban geographers, it is unusual to create urban agglomerations on the basis of transport time. For example, Matthiessen & Schwarz (1999) merge Edinburgh and Glasgow to be a joint 'research centre', but organizations involved in collecting and analyzing spatial data (e.g., the ESPON project being committed to delineating functional urban areas in Europe) consider the two Scottish cities to be separated entities.

In their study, Csomós and Tóth (2016) do not create metropolitan areas for analytical purposes but adopt already existing metropolitan areas as defined by national and international statistical organizations.

The other approach suggests that, in order to avoid problems generated by the non-standardized delineation method of metropolitan areas, the focus should be on cities, more precisely on settlements reported by the authors on the publications as the geographical location of their institutions. From the perspective of the data collection, a 'settlement' corresponds to the spatial unit being indicated on the publication between the name of the country (in some cases the name of the state, region, province, etc.) and the institution. However, in this case a serious problem emerges: many localities of highly heterogeneous sizes will be classified into the same category; for example, Beijing (a megacity), Villejuif, Val-de-Marne (a suburb located 7 km from the centre of Paris), Stanford, California (officially a census-designated location but technically a university campus), Minamitsuru District, Yamanashi (a rural district containing two towns and four villages) and even the Moffett Federal Airfield (a joint civil-military airport being home to the NASA Ames Research Center). As an outcome of this grouping method, the research results will be less comparable and reliable.

In conclusion, one of the fundamental problems of spatial scientometrics focusing on the city level stems from the lack of consensus on how the spatial unit 'city' should be defined and how metropolitan areas should be delineated.

**2.2 Challenges stemming from lack of standardized method of bibliometric data collection on city level**

Even if a consensus on the method regarding the definition and delineation of cities (urban agglomerations, metropolitan areas, etc.) could be achieved among researchers involved in spatial scientometrics, another problem would emerge: there is no standardized method of how bibliometric data should be collected and processed on the city level. In the following section, I demonstrate this problem through the lens of the Web of Science (WoS) database, which is the most frequently employed publication database for bibliometric research. The WoS provides three possible options to obtain bibliometric data on the city level: 1) we can search the name of a city directly (i.e., by simply typing the name of the city in the 'Basic Search' field), 2) we can collect bibliometric data of institutions located in a given city and integrate it on the city level 3) and we can download each publication one by one to extract data from the 'Author Information' field. In the following section, these options are presented through some examples.

1) In the first case, we simply search the name of cities to find out information about their publication characteristics (publication output, the output of highly cited papers, etc.). Let us take an example from the United States by employing the delineation of Metropolitan Statistical Areas (MSAs) and Combined Statistical Areas (CSAs) introduced by the Office of Management and Budget.

In 2016, researchers affiliated with New York City (New York, NY in the WoS) published 39,646 SCI/SSCI papers. New York City is the most populous settlement within the New York metropolitan area, the delineation of which has several approaches but it most commonly corresponds to either the New York-Newark-Jersey City, NY-NJ-PA Metropolitan Statistical Area or the more extensive New York-Newark, NY-NJ-CT-PA Combined Statistical Area (the latter one is also used by the United Nations in its World Urbanization Prospects). Within the MSA (and the CSA either), several types of settlements and 'de facto' settlements are located, such as cities, towns, boroughs, townships (located primarily in the states of New Jersey and Pennsylvania), villages, hamlets (unincorporated settlements in the state of New York) and census-designated places (unincorporated small communities defined by the United States Census Bureau for statistical purposes). In the following analysis, only the settlement types of cities, towns, boroughs and townships are involved because they are the largest ones in terms of population, and they are considered 'de jure' settlements. Within the New York-Newark-Jersey City, NY-NJ-PA Metropolitan Statistical Area (henceforward referred to as New York MSA), 466 such settlements are located out of which, in 2016, 286 were reported by authors on the publications as the geographical location of their affiliation. Furthermore, the New York-Newark, NY-NJ-CT-PA Combined Statistical Area (henceforward referred to as New York CSA) is home to 241 additional settlements, out of which 79 were the publication locality. That is, 708 settlements (i.e., cities, towns, boroughs and townships) are located in the New York CSA (including New York City), out of which 366 were reported by authors on the publications as the geographical location of their affiliation. If we determine the publication output of those 366 settlements one by one then simply summing them up, we find that in 2016, 79,691 SCI/SSCI papers were produced in the New York CSA. The contribution of New York City to the total output of the New York CSA is approximately 50 percent.

However, because the above method is based on 'integer counting', it provides false and misleading information on the real number of papers published in the New York metropolitan area. This problem will be demonstrated through an example. If we simply sum up the publication output of the top 25 publishing localities in the New York CSA, the result will be 73,038 papers (Table 2). By using this method, papers co-authored by researchers located in two or more settlements within the New York CSA will be counted twice or many times. For example, if a paper has an author from New York City, and another from New Brunswick (the fifth largest publishing locality in terms of the number of papers), it will be counted twice by using integer counting; however, on the level of the New York CSA, it should be counted only once. That is, rather than using the integer counting approach, papers should be summed up by fractional counting to attribute credit proportionally in the case of multi-authored papers. This procedure requires a technical adjustment on the WoS search page: the search fields (rows) should be connected by 'OR' and not 'AND'. This method allows us to count multi-authored papers published by authors affiliated with one of the settlements within the New York CSA (or any other metropolitan areas in the world) only once. If we set the logical connection to 'OR' in the case of the top 25 publishing localities of the New York CSA, the result will be 67,642, which is 92.6 percent of the value we have received by using integer counting. The problem is that on the WoS search page, the maximum number of search fields cannot exceed 25; consequently, by applying this method, it is technically not possible to determine the publication output of more than 25 settlements by fractional counting. The solution seems to be particularly time-consuming: each of the 79,691 papers published in the New York CSA in 2016 should be downloaded and each paper should be checked one by one to filter out duplicates.

Table 2. The top 25 settlements located in the New York CSA in terms of SCI/SSCI publication output (2016)

| Rank | Settlement | Settlement type | County | State | Metro* | Number of publications in 2016 |
|---|---|---|---|---|---|---|
| 1 | New York City | City | - | New York | MSA | 39,646 |
| 2 | New Haven | City | New Haven County | Connecticut | CSA | 9,578 |
| 3 | Princeton | Borough | Mercer County | New Jersey | CSA | 5,020 |
| 4 | Rochester | Town | Ulster County | New York | CSA | 4,311 |
| 5 | New Brunswick | City | Middlesex County | New Jersey | MSA | 2,651 |
| 6 | Piscataway | Township | Middlesex County | New Jersey | MSA | 2,357 |
| 7 | Newark | City | Essex County | New Jersey | MSA | 2,215 |
| 8 | Bethlehem | City | Lehigh County | Pennsylvania | CSA | 839 |
| 9 | West Haven | City | New Haven County | Connecticut | CSA | 806 |
| 10 | East Hanover | Township | Morris County | New Jersey | MSA | 722 |
| 11 | Kenilworth | Borough | Union County | New Jersey | MSA | 722 |
| 12 | Hempstead | Town | Nassau County | New York | MSA | 412 |
| 13 | Orange | Township | Essex County | New Jersey | MSA | 341 |
| 14 | Hyde Park | Town | Dutchess County | New York | MSA | 338 |
| 15 | Hoboken | City | Hudson County | New Jersey | MSA | 326 |
| 16 | Hackensack | City | Bergen County | New Jersey | MSA | 318 |
| 17 | Raritan | Borough | Somerset County | New Jersey | MSA | 313 |
| 18 | Summit | City | Union County | New Jersey | MSA | 311 |
| 19 | Rahway | City | Union County | New Jersey | MSA | 299 |
| 20 | Montclair | Township | Essex County | New Jersey | MSA | 298 |
| 21 | Allentown | City | Lehigh County | Pennsylvania | CSA | 298 |
| 22 | Bridgewater | Township | Somerset County | New Jersey | MSA | 269 |
| 23 | Ridgefield | Town | Fairfield County | Connecticut | CSA | 220 |
| 24 | White Plains | City | Westchester County | New York | MSA | 216 |
| 25 | Morristown | Town | Morris County | New Jersey | MSA | 212 |
| | Integer counting | | | | | 73,038 |
| | Fractional counting | | | | | 67,642 |

The problem stemming from the use of different counting methods when examining the publication output of cities vs. metropolitan areas seems serious, but it appears to be more problematic when the scientometric analysis focuses on exploring the collaboration between two (or more) metropolitan areas. In this case, settlements should be organized into a matrix of metropolitan area dyads, and the collaboration between each element (i.e., publishing locality) of the matrix should be examined one by one. For example, Upton, NY (home to the Brookhaven National Laboratory) produced 1,335 papers in 2016, while authors affiliated with Berkeley, CA (home to the University of California, Berkeley and the Lawrence Berkeley National Laboratory) published 9,764 in the same year. In 2016, a total number of 228 papers were published with the contribution of authors from both Upton and Berkeley. Upton is part of the New York CSA, a metropolitan area containing more than 1,000 settlements, and Berkeley is located in the San Jose–San Francisco–Oakland, CA Combined Statistical Area, having approximately the same size in terms of the number of settlements as the New York CSA. That is, the collaboration of a given settlement located in a given metropolitan area with settlements located vis-à-vis in the other metropolitan area should be checked one by one, and in the case of two extensive metropolitan areas containing hundreds of settlements with publication data, the number of collaboration links can range from 10,000 to 100,000 or more.

2) The publication output of a city (metropolitan area) can also be investigated through the lens of institutions located in that city (metropolitan area). Practically, this means that the bibliometric data of institutions should be integrated on the city level (see, for example, Bornmann and de Moya-Anegón 2018, 2019). It is a serious advantage of this approach that through the InCites, a customized, citation-based research analytics tool, the WoS provides significantly more accurate bibliometric data on institutions as compared to cities. However, we have to face a general disadvantage stemming from the problematic use of integer vs. fractional counting approaches. For example, the SCImago Institutions Rankings (SIR)[5] (used by Bornmann and de Moya-Anegón (2018) and Gómez-Núñez et al. (2016) in their studies classified 31 institutions into New York City, an additional 28 institutions into settlements of the New York MSA and 14 institutions into settlements located in the New York CSA but outside the New York MSA. That is, based on information provided by the SIR, 73 top research institutions are located in the New York CSA, and most possibly this number is only a fraction of the real number of research institutions situated in the New York CSA. Out of the New York CSA-based 73 institutions, the top 25 institutions published 63,692 SCI/SSCI papers measured by integer counting, but the number of papers was only 51,784 if using fractional counting (i.e., 18.7 percent of all papers were co-authored by researchers affiliated with institutions located in different settlements within the New York CSA).

When focusing on integrating bibliometric data of institutions on the city level or the metropolitan area level, we have to face another problem. Many institutions (e.g., universities and major governmental research institutions) and companies in the world are considered highly complex organizations having faculties, research centres and subsidiaries located outside the city or the metropolitan area where the given institution or company is headquartered. For example, IBM is headquartered in Armonk, New York, a 'hamlet' located in the New York MSA. In the InCites, all papers authored by researchers affiliated with IBM are simply assigned to IBM. Does it mean that those papers have been produced in Armonk? In 2016, IBM produced 876 SCI/SSCI papers, out of which only 30 papers (3.4 percent) were contributed by Armonk-based researchers. IBM (just like other multinational companies) maintains a spatially complex structure and publications can come from many cities located in (or outside of) the United States where IBM operates subsidiaries (e.g., San Diego, California and San Jose, California) (see Csomós and Tóth 2016). This problem also emerges in the case of multi-campus universities located particularly in Australia (Scott et al. 2007), the United States and Western Europe (Pinheiro and Nordstrand Berg 2017). For example, authors located in Creswick, Victoria, Australia and affiliated with the School of Ecosystem and Forest Sciences of the University of Melbourne contributed 49 publications in 2016. Creswick is located approximately 130 kilometres from Melbourne, and it is definitely not part of Greater Melbourne; yet, papers authored by Creswick-based researchers are assigned to the University of Melbourne by InCites as if they were produced in Melbourne.

That is, publications being assigned to an institution by the InCites (or the WoS) do not necessarily mean that those publications are produced in the locality where the institution is headquartered; consequently, each publication assigned to a given institution should be checked one by one to extract accurate geographical data from the author information field.

---

[5] The SCImago Institutions Rankings (SIR) is a classification of academic and research-related institutions ranked by a composite indicator that combines three different sets of indicators based on research performance, innovation outputs and societal impact measured by their web visibility.

Finally, the aforementioned approaches would be significantly more complicated if the fractional counting method was applied not only when investigating the publication output of cities (by summing up the output of institutions) and metropolitan areas (by summing up the output of settlements) but focusing on the publication itself. Most publications are multi-authored, and co-authors can be located in many cities in the world. In the case of multi-authored papers, the credit should be attribute proportionally to co-authors. For example, if a paper is authored by researchers located in Tokyo and Beijing, the proportion of each city in the collaboration account for 1/2. That is, the fractional counting approach would be required to use not only vertically (when we integrate bibliometric data of institutions and/or cities into the metropolitan area level) but also horizontally (when we determine the credit of a given author to the contribution).

3) By accepting the critical observations of Gauffriau et al. (2008) regarding the use of fractional counting in bibliometrics, the most accurate way to define the publication output, etc. of metropolitan areas would be based on the use of fractional counting approach. However, as can be seen in Table 1, there is only one research group that measures publication output of metropolitan areas (referring to them as urban agglomerations) by using fractional counting (Grossetti et al. 2014; Maisonobe et al. 2016, 2017, 2018b, 2019). The reason for this issue is that the process of data extraction from the datasets provided by the WoS cannot be automatized (for example, only 500 records are allowed to be downloaded from the WoS at once, whereas some cities produce more than 100,000 papers per year), and it may take years for even a research team with several members to compile a dataset that can serve special analytical purposes[6].

## 2.3 Challenges stemming from the lack of a clearly defined concept of what city-level scientometric analysis serve

In the previous sections, the focus was on methodological challenges stemming from the lack of standardized methods of how cities (more precisely: metropolitan areas) should be delineated and how bibliometric data on the city level should be collected and processed. However, no matter how hard the above problems appear, eventually it is (or rather it would be) possible to bridge them over. There are some more crucial questions that need to be answered: What do we intend to demonstrate through bibliometric data of cities? What is the main goal of conducting scientometric analysis focusing on the city level? Most studies, primarily those being produced by experts in the field of bibliometrics, attribute high importance to methodological issues without providing an adequate answer regarding the practical goal of the research. Naturally, a newly introduced method that significantly contributes to the development of the field of spatial scientometrics can be of high importance, but it remains uncovered how a city can profit from the results. Geographers like to observe cities as elements of networks in which their hierarchical positions change over time and bibliometric data of cities are perfectly suitable to demonstrate changes (and spatial biases). According to Matthiessen and Schwarz (1999), 'in a world where the importance of regions and cities as sites of competition and producers of strategic plans is increasing', it is of significant interest to identify the comparative position of cities in science based on their scientific strengths (measured in terms of publication output). In conclusion, it seems that it is substantially important to determine the scientific strength, publication output, collaboration pattern, etc. of cities (by using bibliometric and scientometric methods developed by experts) because, based on the information they provide, cities will be able to react and change their position in science by incorporating modifications in the science system they host. This is, however, not so evident.

---

[6] It should be noted that some research organizations, specialized to produce bibliometric and/or scientometric analysis for both professionals and academics, have special agreements with Clarivate Analytics, the owner and operator of the Web of Science, which allows them to obtain pre-processed data from the in-house version of the WoS. Amongst these organizations, there are the Centre for Science and Technology Studies (CWTS) from Leiden and the Montreal-based L'Observatoire des sciences et des technologies (OST).

This problem has been generated by the fact that institutions located in a given city are part of science system being operated and managed independently from the city, and the city cannot control the operation of the local elements of that system. Generally, they are the national governments (in some cases the governments of federal states, provinces, etc.), and in the case of the Europe Union, the Commission, who have the legal authorization and power to shape the national (and the supranational) science system through regulations and funding, and who have direct impact on the operation of the system and indirect on the outcome (Celis and Gago 2014; Choi et al. 2009; Jacob and Lefgren 2011; Lemola 2002; Lepori et al. 2007; Ling and Naughton 2016; van der Meulen 1998). If a specific discipline is donated more money by the government, it will generate more research activity in that field and will eventually be resulted in the production of more publications and/or patents (Moed et al. 1998; Mongeon et al. 2016; Wang and Shapira 2015). Institutions (universities, governmental research institutions, even companies) are also capable of controlling the research activity; moreover, as employers, they can establish requirements against researchers whom they are contracted with (by, for example, specifying the expected outcome of a research project). In addition, almost every university has a well-defined research strategy indicating a straightforward message for the faculty and the public as well: the university goal is to become a top 100 university, the university will be the leading research institution in the continent by…, the university will increase the number of international students, the university's main goal is to acquire more research grants, and the university will significantly increase its publication output and impact, etc. Whether observing countries or institutions, they have well-defined policy goals being supported by both regulations and financial instruments, and all participants of the science system can be identified (i.e., in the case of countries they are universities, research institutions, etc., and in the case of institutions they are individual researchers). In the case of countries and institutions, it is possible to monitor whether the target indicators defined in the strategies have been fulfilled by a certain date and if not, what interventions are necessary to address to correct them. In contrast, most cities in the world do not have science policy, and they cannot control the operation of institutions they host. Bibliometric and scientometric data on the city level cannot provide information on the performance of the science system located in cities because cities are rather considered to be containers of the elements of the national or international science system operating independently from cities. Naturally, some exceptions can be found in the world. For example, within the organization of the Government of Beijing Municipality, there is the Beijing Municipal Commission of Science and Technology, the duty of which is to 'make policies and measures to reform the municipal S&T system' [7]. It should be noted, however, that Beijing is not only a city in the usual sense of the term but is a province as well, a first-tier division in the Chinese administrative system.

In addition, another phenomenon emerges making it quite problematic to observe institution located in a given city as elements of a local science system. Bornmann and de Moya-Anegón (2018) notice that 'institutions are frequently spatially clustered in larger cities'. According to Catini et al. (2015) 'research institutions involved in scientific and technological production' generally locate close to each other and produce well-outlined research clusters within cities. It seems logical that the geographically clustered research institutions, universities and companies establish intense research collaboration which will be reflected in a large number of co-authored papers. This phenomenon should, however, be approached more carefully.

For example, Geneva, Switzerland is home to three world-renowned research institutions producing a large number of scientific publications year by year, which are as follows: the European

---

[7] Beijing Municipal Commission of Science and Technology
http://www.ebeijing.gov.cn/Government/Departments/t930030.htm

Organization for Nuclear Research (CERN)[8], the University of Geneva and the World Health Organization (WHO). In 2016, authors affiliated with CERN published 1,216 SCI/SSCI papers, the output of the University of Geneva was 4,406 papers, while the Geneva-based WHO produced 1,004 papers. CERN is located eight kilometres from both the University of Geneva and the WHO, while the distance between the University of Geneva and the WHO is less than three kilometres. However, only CERN and the University of Geneva produced a less significant collaboration by publishing 340 co-authored papers in 2016. Hence, based on the total number of papers written in collaboration with authors affiliated with other institutions, CERN was ranked seventh among the top collaborators of the University of Geneva, and for CERN, the University of Geneva was only the 14$^{th}$ collaborator. In 2016, the University of Geneva and the WTO published not more than 45 co-authored papers, while, at least based on the number of co-authored papers, CERN and the WTO did not collaborate with each other. That is, despite the fact that three prestigious research institutions regularly producing a high publication output are located in the same city, within an area of ten square kilometres (it can therefore be stated that they are geographically clustered), the collaboration intensity between them is quite low (much lower than would be expected)[9]. In the case of other cities worldwide, a similar pattern can be identified and the bigger the city is, the less intense intraurban collaborations emerge between local institutions.

## 3 Discussion and conclusion

Challenges on spatial bibliometrics and scientometrics focusing on the city level stem from three main problems: 1) there are no standardized methods of how cities should be defined and metropolitan areas should be delineated, 2) there is no standardized method of how bibliometric data on the city level should be collected and processed and 3) it is rather unclear regarding what city level scientometric analysis should demonstrate and justify. Because the aforementioned problems are characterized by a mixture of methodological and theoretical issues, the solutions require complex approaches.

    1) The definition of the spatial unit 'city' is difficult and sometimes based on subjective intuitions. In the literature, the term 'city' can correspond to a range of spatial units including cities (as defined by national administrative systems), urban agglomerations, metropolitan areas, metropolitan regions and city-regions. Furthermore, the analytical purpose significantly influences which of the 'city' categories should be taken into consideration. In scientometric studies, the most commonly used spatial categories are administratively defined cities, and metropolitan areas delineated for special analytical purposes. Most probably, the challenges stemming from the varying definitions of cities and the non-standardized delineation method of metropolitan areas will always remain a problematic issue in spatial scientometrics, just like in the case of urban geography, which has been focusing on this research topic for decades. To avoid problems due to the lack of a standardized delineation method of metropolitan areas, it would be an ideal solution to accept the metropolitan area definitions of national and/or international statistical organizations. It may be that these approaches are not considered to be the best option for special analytical purposes (see, Matthiessen and Schwarz 1999; and Maisonobe et al. 2018b), but at least their use in spatial scientometrics can lead to consensus among researchers and can make research outcomes more comparable.

---

[8] More precisely, the main site of the CERN is in Meyrin, a commuter town (a small village formerly) located eight kilometers from Geneva. However, in 95 percent of all cases, authors indicate Geneva and not Meyrin on the publications as the geographical location of their affiliation.

[9] The main reason why CERN, the University of Geneva and the WHO produce low collaboration, despite being located geographically close to each other, is that they are characterized by different disciplinary profiles. The WHO regularly publishes a large number of papers in the field of Public, Environmental and Occupational Health, the main research areas of the University of Geneva belong to the fields of physics (producing 340 co-authored papers with CERN) and medicine, whereas most papers coming from CERN have been written in particle physics and other subfields within physics. When establishing scientific collaboration, the most significant factor is supposed to be the disciplinary and not the geographical proximity.

2) It is a more problematic question regarding how to obtain appropriate bibliometric and scientometric data for spatial analysis. The problem is generated by the fact that the outcomes of spatial analyses are significantly influenced by the quality of raw bibliometric data provided by publication databases. The most frequently used database in spatial scientometrics is the Web of Science (primarily the SCI and SSCI databases), the analytical tool of which — InCites — provides bibliometric data on country (in some cases regional), institutional and individual levels but neglects to provide data for the city level. Hence, we can only obtain accurate bibliometric data on cities if we manually scrutinize each paper one by one to extract information on the geographical location of authors' affiliation. Having no possibility to automatize the process, it seems particularly complicated to examine, for example, 500 papers (i.e., the maximum number of records the WoS allows us to download at once), and it is almost impossible to scrutinize tens of thousands or millions of papers one by one (unless we have years to do it). To avoid this problem, most researchers use the integer counting approach for scientometric analysis on the city (and metropolitan area) level because it does not require special and time-consuming data processing methods. Naturally, the most optimal solution would be if the WoS provided pre-processed bibliometric data on the city level through the InCites platform as it does in the case of countries. By obtaining standardized city level bibliometric data, it would become much easier to calculate the publication output, etc. of metropolitan areas by using fractional counting approaches.

3) Finally, we should define the exact and practical goal of bibliometrics and scientometrics focusing on the city level. The problem comes from the fact that institutions (universities, governmental and corporate research institutions, etc.) located in a given city are independent entities from the city (more precisely: the local government) and they are part of a larger (national and/or international) science system. In addition, while it is observed that research institutions tend to locate in geographical clusters within cities, they only hypothetically produce strong scientific collaboration networks. But the truth is that cities benefit of being home to universities, research institutions, large hospitals and R&D-driven and innovative firms because these organizations attract members of the creative class (Van Noorden 2010). This class is a fast-growing, highly educated and well-paid segment of the workforce whose presence positively affects the development of the local economy and the general performance of the city and has positive impacts on labour productivity (Florida et al. 2008). In the hinterlands of research universities, new innovative start-up firms are established offering jobs for talented and highly educated people. Due to these developments, the city is required to enhance the quality of public services and develop appropriate infrastructure. This self-reinforcing process gives significant impetus to the growth of the creative class and finally transforms the city into a thrilling place being characterized by innovative and creative milieu. Boston is considered the best example to demonstrate this socioeconomic process (Owen-Smith and Powell 2004; Tödtling 1994). This mechanism is also underpinned by the fact that in 2017 General Electric (GE), one of the largest industrial conglomerate of the United States, decided to relocate its headquarters from Fairfield, Connecticut to Boston, being described as a 'dynamic and creative city' by Jeff Immelt, Chairman and CEO of the company[10]. The relocation of GE's headquarters to Boston has been motivated by the following main factors: the Greater Boston area is home to 55 colleges and universities (including Harvard University and MIT), Massachusetts spends more on research and development than any other region in the world, and Boston attracts a diverse, technologically fluent workforce focused on solving challenges for the world (Florida 2016). As an outcome of this development, we can predict that R&D activity will increase in Boston that may result in a larger scientific output in terms of the number of publications and patents. By using bibliometric and scientometric methods, we can measure and evaluate how the scientific and innovative potential of a city change over time. That is, indicators related to the scientific performance and characteristics of institutions located in a given city are indirectly reflected in the city itself.

---

[10] Announcement of the GE: GE Moves Headquarters to Boston. https://www.genewsroom.com/press-releases/ge-moves-headquarters-boston-282587

At this point, one more question arises: If we accept the fact that elements of the science system located in a city operate quasi independently from the city, what possibilities does the local government have to positively change the aggregate value of such science indicators. We may think that cities have highly limited opportunities to compete for components of the science establishment because most of them (especially universities) are strongly tied to their original locus. However, if cities can create vibrant and smart environments with high-quality public services and infrastructure, they can turn themselves into magnets for attracting innovative firms, and governmental and corporate research centres. These factors were of significant importance when GE decided to move to Boston, and they were also highly important when the consortium of member countries chose Lund, Sweden (against Bilbao, Spain, and Debrecen, Hungary) to be the home for the European Spallation Source (ESS), the world's most powerful pulsed neutron source.

In conclusion, there are some major challenges ahead of spatial scientometrics focusing on the city level, but the necessity of this kind of research is clear and well underpinned.